\documentclass[pra,aps,twocolumn,superscriptaddress]{revtex4}
\usepackage{graphicx}
\usepackage{amsmath}

\newcommand{\eq}{\begin{equation}}
\newcommand{\fine}{\end{equation}}

\begin{document}

\title{
Baryons made of heavy quarks at the LHC}
\author{Andr\'e Martin }                                       
\email{martina@mail.cern.ch} 
\affiliation{Theoretical Physics Division,CERN, Geneva}
\begin{abstract}
Baryons made of heavy quarks are extremely interesting and could be seen at the LHC.
\end{abstract}
\maketitle

        This is a short report on baryons made of heavy quarks,
i.e. t, b, c and s, a work which was initiated in discussions with
Jean-Marc Richard and Tai Tsun Wu. It happens that nothing of this work
was ever published, except for the fact that Jean-Marc Richard touched
that subject in a recent talk at a workshop organized by LHCb. So, I
decided to write something myself, though I owe a lot to Jean-Marc and
Tai.

        The extraordinarily large luminosity of the LHC makes it
possible to see particle systems whose creation is not very likely and
which are extremely interesting.
  To begin I start with the fact that the belief that the top quark
never hadronizes can be challenged. Indeed, out of 20 million top
quarks, 1000 of them have life which is 10 times the official lifetime and
have time to hadronize. How? I don't want to fight with my collegues, but
if one could see, for instance, tbb it would be fantastic!
    To be more realistic we concentrate on baryons made of b, c, and s
quarks.  In particular the dream of Bj\"orken to see a ccc baryon becomes
accessible, but you can see more than that, like the bbb baryon. The
masses of these particles can be calculated using a potential model
proposed by Jean-Marc Richard in 1981 \cite{Richard1981}, inspired by a potential that 
I had proposed  earlier for mesons \cite{Martin1980}. These potentials, even though
there is hardly a good justification for them, are extremely successful.
for instance they give tha mass of the $\Omega^{-}$ with an accuracy of
0.2\% or the mass of the $B_{s}$ with an accuracy such that when the late
Lorenzo Fo\`a announced its dicovery by Aleph he said that he did not need
to give tha value of the mass since I had predicted it already. As
Rudolf Faustov noticed you dont need to use more sophisticaed methods to
get these masses. 

 What seems to me extremely interesting is the chain decays of
these baryons. You go from one baryon to the next by emitting either a
lepton pair or a meson with a change of charge of one unit, and,
provided the initial baryons are produced with sufficient energy, you
can see the various steps because of the relatively large lifetime of
the b quarks of $1.6\times10^{-12}$ secs. Let me take one example,
which is probably  not the most likely, but pedagogically simple. You
can have

             bbb (charge -1) $\rightarrow $ bbu (charge 0, invisible) $\rightarrow $ buu
(charge +1) $\rightarrow $ uuu (i.e. $\Delta^{++}$, charge +2) $\rightarrow $ uud (proton) 

     On all the visible elements of the chain, charge, mass, lifetime
can be measured. The above potential model can be tested. Branching
ratios can improve the information on the CKM matrix.
Seen by a perhaps na\"{i}ve theoretician, this seems  very promising.
Marek Karliner informed me that he and Jonathan Rosner have written a 
paper on Baryons containing 2 heavy quarks \cite{Karliner2014}.

\end{document}